\documentclass[12pt]{article}
\usepackage{amsfonts,amssymb,enumerate,epsf}
 \usepackage[english]{babel}
\newtheorem{theorem}{Theorem}

\newtheorem{lemma}[theorem]{Lemma}
\newtheorem{proposition}[theorem]{Proposition}

\newtheorem{remark}[theorem]{Remark}

\begin{document}

\title{On ground state of some non local Schr\"{o}dinger operators.
\thanks{The work is partially supported by SFB 701 (Universitat Bielefeld).
The research of Sergey Pirogov and Elena Zhizhina (sections 2-3)
was supported by the Russian Science Foundation, Project ¹
14-50-00150, Stanislav Molchanov was supported by NSF – grant DMS – 1008132 (USA). } }

\author{Yuri Kondratiev\thanks{Fakultat fur
Mathematik, Universitat Bielefeld, 33615 Bielefeld, Germany
(kondrat@math.uni-bielefeld.de).} \and Stanislav Molchanov\thanks{Department of Mathematics and Statistics, UNC Charlotte(USA) and University Higher School of Economics, Russian Federation (smolchan@uncc.edu)}
\and Sergey Pirogov\thanks{ Institute for Information Transmission
Problems, Moscow, Russia (s.a.pirogov@bk.ru).} \and Elena
Zhizhina\thanks{Institute for Information Transmission Problems,
Moscow, Russia (ejj@iitp.ru).} }

\date{}

\maketitle

\begin{abstract}
We study a ground state of some non local Schr\"{o}dinger operator
associated with an evolution equation
for the density of population in the stochastic contact model in
continuum with inhomogeneous mortality rates.
We found a new effect in this model, when even in the high dimensional
case the existence of a small
positive perturbation of a special form (so-called, small paradise)
implies the appearance of the ground state.
We consider the problem in the Banach space of bounded continuous
functions $C_b (R^d)$ and in the Hilbert space $L^2(R^d)$.



\end{abstract}

\section{Introduction}

The asymptotic behavior of stochastic infinite-particle systems in continuum
can be studied in terms of evolution equations for correlation functions. For the stochastic contact model in the continuum \cite{KS} the evolution equation for the first correlation function
(i.e.,  the density of the system) is closed and it can be considered separately from equations for higher-order correlation functions \cite{KKP}. In this case we have the following evolution problem for $u~\in~C( [0,\infty); {\cal E} )$ associated with a nonlocal diffusion generator $L$:
\begin{equation}\label{PP}
\frac{\partial u}{\partial t} \ = \ L u, \quad u = u (t, x), \quad x \in \mathbb{R}^d, t \ge 0, \quad u(0,x) = u_0 (x) \ge 0.
\end{equation}
in a proper functional space ${\cal E}$. As ${\cal E}$, we consider in our paper two spaces: $C_b ({\mathbb R}^d)$, the Banach space of bounded continuous functions on $\mathbb{R}^d$, and $L^2 ({\mathbb R}^d)$. These spaces are corresponding to two different regimes in the contact model:  systems with bounded density and ones essentially
localized in the space.

The operator $L$ has the following form:
\begin{equation}\label{L}
Lu(x) \ = \  - m(x) u(x) \ + \ \int_{\mathbb R^d} a(x-y) u(y) dy,
\end{equation}
where $a(x) \ge 0, \; a \in L^1({\mathbb R}^d)$ is an even continuous function such that:
\begin{equation}\label{a1}
\int_{\mathbb R^d} a(x) dx = 1.
\end{equation}
Since $a(x)$ is bounded, then $a(x) \in L^2 ({\mathbb R}^d)$ and
\begin{equation}\label{a4}
\tilde a(p) = \int_{\mathbb R^d} e^{-i(p, x)} a(x) dx \in L^2 ({\mathbb R}^d) \quad \mbox{with  } \quad |\tilde a(p)| \le 1.
\end{equation}
The function $a(x-y)$ is the dispersal kernel associated with birth rates in the contact model.
The function $m(x)$ is related with mortality rates. We assume here that
\begin{equation}\label{mm}
m(x) \in C_b ({\mathbb R}^d), \quad  0 \le m(x) \le 1, \quad m(x) \to 1, \; |x| \to \infty.
\end{equation}

The contact model with  homogeneous mortality rates $m(x) \equiv const$ has been studied in \cite{KKP}. It was proved there that only in the case $m(x) \equiv 1$ there exists a family of stationary measures of the model (for $d \ge 3$); if $m(x) \equiv m \neq 1$, then the density of population either exponentially growing (supercritical regime: $m<1$) or exponentially decaying (subcritical regime: $m>1$).

Here we are interesting in local perturbations of the stationary regime, when $m(x)$ is an inhomogeneous
in space non-negative function. We prove that local fluctuations of the mortality with respect to the critical value $m(x) \equiv 1$ can push the system away from the stationary regime. As a result of such local perturbations, we will observe exponentially increasing density of population everywhere in the space.
The goal of this paper is to obtain conditions on the mortality rates which give the existence of a positive discrete spectrum of the operator (\ref{L}) in the spaces $C_b ({\mathbb R}^d)$ and $L^2 ({\mathbb R}^d)$, and to prove the existence and uniqueness of a positive eigenfunction $\psi(x)>0$ corresponding to the maximal eigenvalue $\lambda_0>0$ of the operator $L$. We call this function the ground state of the operator $L$.

The existence of not only the principal eigenvalue $\lambda_0$ but also the corresponding positive eigenfunction $\psi$ is very important in the analysis of the long-time behavior of non-local evolution problems. The so-called principal eigenpair problem $(\lambda_0, \psi)$ for some nonlocal dispersal operators has been studied recently in many papers, see e.g. \cite{Cov, BC} and references therein.  However the results of these papers were obtained under quite restrictive conditions, namely the assumption that the dispersal kernel $a(x-y)$ is a function with a compact support, or under the condition that the eigenvalue problem is studied in a bounded domain $\Omega \subset R^d$. We consider in our paper the case of unbounded domain $\Omega = R^d$, and the only assumption on the dispersal kernel is the integrability condition.

We prove in our paper that the ground state appears in two cases: either there exists such a region of any (small) positive volume, where the fluctuation $V(x)$ is equal to 1 (Theorem \ref{Theorem 2}), or $V(x)$ is positive and less than 1 in a large enough region (Theorem \ref{Theorem 3}). We stress that the function $V(x)$ should be bounded from above by 1, since the mortality $m(x) \ge 0$ is a non-negative function. Thus we observe for the nonlocal operator (\ref{L}) new effects different from those of the Shr\"{o}dinger operators. We found that in the high dimensional case some small (in the integral sense) perturbations $V(x)$ of the mortality $m(x)$ from the critical value $m(x) \equiv 1$ imply existence of the positive eigenvalue and the corresponding positive eigenfunction. In small dimensions $d=1,2$, the ground state of the operator $L$ exists for any small local positive fluctuation $V(x)=1-m(x)$ of the mortality $m(x)$ from the critical value under condition that the dispersal kernel has the second moment (Theorem \ref{Theorem 1}). We also proved the analogous results for the model in the subcritical regime (Theorem \ref{Theorem 12}).

\section{Spectral properties of $L$}

In this section we describe a general approach to study a discrete spectrum of the operator $L$ in both spaces $C_b ({\mathbb R}^d)$ and $L^2 ({\mathbb R}^d)$. This approach is based on the analytic Fredholm theorem and the study of a spectral radius of compact operators.

The operator $L$ can be rewritten as
\begin{equation}\label{2}
L u(x) \ = \  L_0 u(x) \ + \ V(x) u(x), \quad u(x) \in  C_b({\mathbb R}^d),
\end{equation}
$$
L_0 u(x) \ = \ \int_{\mathbb{R}^d} a(x-y) (u(y) - u(x)) dy, \quad  V(x) \ = \
1-m(x).
$$
The potential $0 \le V(x) \le 1$ describes local (negative) fluctuations of the mortality, and we assume that
\begin{equation}\label{6A}
V(x) \in C_b({\mathbb R}^d) \quad \mbox{ and }  \quad \lim_{|x| \to \infty} V(x) = 0.
\end{equation}

The operator $L_0$ is bounded and dissipative in $C_b({\mathbb R}^d)$:
$$
\| (\lambda - L_0)f \| \  \ge \  \lambda \|f\|  \quad \mbox{ for } \quad \lambda \ge 0.
$$


\begin{lemma} \label{Lemma 1}
 The operator $L$ has only discrete spectrum in the half-plane $$\mathcal{D} = \{ \lambda \in \mathbb{C} \ | \ \ Re \lambda>0 \}.$$
\end{lemma}

{\it Proof} is based on the analytic Fredholm theorem for analytic operator-valued functions, see \cite{DS}, \cite{RS1}. Since for any $\lambda \in \mathcal{D}$  the resolvent $(L_0 - \lambda)^{-1}$ is a bounded operator, we have
\begin{equation}\label{5A}
(\lambda  - L_0 - V)= (\lambda  - L_0) (1 - (\lambda  - L_0)^{-1} V),
\end{equation}
and
\begin{equation}\label{1L}
(\lambda  - L_0 - V)^{-1} = (1 - (\lambda  - L_0)^{-1} V)^{-1} \ (\lambda  - L_0)^{-1}.
\end{equation}
Using the Neumann series for the operator $(\lambda  - L_0)^{-1}$ we get:
\begin{equation}\label{mu}
(\lambda - L_0)^{-1} \ = \ \frac{1}{\lambda+1} \ + \ \frac{1}{\lambda+1} A_{\lambda},
\end{equation}
where
\begin{equation}\label{a}
A_{\lambda} = (\lambda+1) (\lambda - L_0)^{-1} - 1 =
\sum_{n=1}^{\infty} \frac{a^{\ast n}}{ (\lambda+1)^n},
\end{equation}
and $A_\lambda$ is a bounded convolution operator when $\lambda \in \mathcal{D}$. The decomposition (\ref{mu}) for $(\lambda -
L_0)^{-1}$ is the crucial point of our reasoning. The kernel of $A_{\lambda}$ is
\begin{equation}\label{G}
G_{\lambda}(x-y) =
\left(  \sum_{n=1}^{\infty} \frac{a^{\ast n}}{ (\lambda+1)^n} \right) (x-y)  =  \frac{1}{(2 \pi)^d} \int_{\mathbb{R}^d} e^{-i(p, x-y)} \frac{\tilde a (p) \ d p}{\lambda + 1- \tilde a(p)},
\end{equation}
with
\begin{equation}\label{10A}
\tilde G_{\lambda}(p) = \frac{\tilde a (p)}{\lambda + 1- \tilde a(p)} \in  L^2({\mathbb R}^d) \quad \mbox{ when } \;  \lambda \in \mathcal{D},
\end{equation}
and, in particular,
\begin{equation}\label{D}
G_{\lambda}(u) \in  C_b({\mathbb R}^d) \cap  L^2({\mathbb R}^d) \quad \mbox{ and } \quad \int_{\mathbb{R}^{d}} G_{\lambda}(u) \ du \ < \  \infty, \quad \forall \lambda \in \mathcal{D}.
\end{equation}
We denote by $W_\lambda$ an operator of multiplication by the function
$$
W_\lambda = 1-\frac{V}{\lambda+1}.
$$
It is a bounded operator with the bounded inverse operator $W_\lambda^{-1}$ when $\lambda \in \mathcal{D}$. Then (\ref{mu}) implies
$$
1 - (\lambda - L_0)^{-1} V =  W_\lambda - \frac{1}{\lambda+1} A_\lambda V,
$$
and we can rewrite (\ref{1L}) in the following way:
\begin{equation}\label{2L}
(\lambda - L_0 - V)^{-1} = ( 1 - Q_{\lambda})^{-1}  \ \left((\lambda - L_0)W_\lambda \right)^{-1},
\end{equation}
where
\begin{equation}\label{Q}
Q_{\lambda} = \frac{1}{\lambda+1} W_\lambda^{-1} A_\lambda V.
\end{equation}

The relations (\ref{G}) - (\ref{D}) for the kernel $G_\lambda (x-y)$ of the operator $A_\lambda$ and the conditions on the potential
$V(x)$ imply that $Q_{\lambda}$ is the compact operator. Really, let us define the truncated operator of multiplication on the potential $V^{(r)}(x) = \chi_r(x) V(x)$ and the convolution operator $A_{\lambda}^{(r)}$ with the truncated kernel $ G_\lambda^{(r)}(x)  = \chi_r (x) G_\lambda(x)$, where $\chi_r(x)$ is the indicator function of the ball $B_r = \{ x: \ |x| <r \}$. Obviously, the operators $ A_\lambda^{(r)} V^{(r)}$ are compact for any $r>0$, and $A_\lambda^{(r)}$, $V^{(r)}$ converge in norm to $A_\lambda$, $V$ correspondingly. Thus we conclude that $A_\lambda V$ is the compact operator.

Consequently,  $Q_{\lambda}$ is an analytic operator-valued function, such that  $Q_{\lambda}$ is a compact operator for any $\lambda \in \mathcal{D}$. Then using the analytic Fredholm theorem \cite{RS1} (Theorem VI.14) we get that the function $(1 - Q_{\lambda})^{-1}$ is a meromorphic function in $\mathcal{D}$. Since $((\lambda  - L_0) W_\lambda)^{-1}$ is a bounded operator, we can conclude that the operator $L$ has only a discrete spectrum in $\mathcal{D}$.
\hfill     $\blacksquare$

\begin{remark}
 \label{Remark 1}
  The representations (\ref{a})-(\ref{G}) imply that \\
1)  $A_{\lambda}$ is a positivity improving operator for all $\lambda>0$, since
$G_{\lambda}(x-y)~>~0, \ \forall \ x,y \in \mathbb{R}^d$,  \\
2)  $G_{\lambda}(x-y) $ is monotonically decreasing with respect to $\lambda>0$, \\
3) formulas (\ref{Q}), (\ref{a}), (\ref{G}) imply that $Q_\lambda, \  \lambda >0,$ is a positivity improving compact integral operator in $C_b({\mathbb R}^d)$ with the kernel
\begin{equation}\label{12A}
{\cal Q}_{\lambda} (x,y) \ = \ \frac{ G_{\lambda}(x-y) V(y)}{ \lambda + 1- V(x)}.
\end{equation}
\end{remark}

We study next the behavior of the spectral radius $r(Q_\lambda)$ of the operator $Q_\lambda$ as a function of $\lambda$
when $\lambda>0$.

\begin{remark}
\label{Remark 2} From the known formula for the spectral radius it follows that if $Q$ is a positive operator, and if there exists a function $\varphi(x) \in {\cal E}$, $\varphi \ge 0$, $\| \varphi \| =1$,
such that
\begin{equation}\label{c0}
Q \varphi (x) \  \ge \ c_0 \varphi(x),
\end{equation}
then
$$
r(Q) \ \ge \ c_0,
$$
(see  e.g. \cite{KR}, Theorem 6.2)
\end{remark}

\begin{lemma}
\label{Lemma 2} The spectral radius $r( Q_{\lambda})$ is continuous and
monotonically decreasing with respect to $\lambda >0$. Moreover, $r(Q_\lambda) \to 0$ for $\lambda \to +\infty$.
\end{lemma}

 {\it Proof.}  The continuity of the spectral radius follows from the compactness and continuity in $\lambda$ of the operator $Q_{\lambda}$.

The second statement of the lemma follows from the fact, that if $A,B$ are positive operators, such that $ A \leq B$ in the order sense defined by the cone of positive functions, then $r(A) \leq r(B)$.
Really, it is easy to see that for both our spaces ${\cal E}$ norms of positive operators $A, B$ are attained on non negative functions:
$$
\| A  \| \ =  \  \sup_{f \ge 0} \frac{\|Af\|}{\|f \|},
$$
and the same for $B$. Thus, $\|A\| \le \|B\|$.

Analogously, for any $n \in N$ we have $\| A^n \| \le  \|B^n \|$, and consequently, $r(A) \le r(B)$ due to the formula
for the spectral radius $r(A) = \lim_{n \to \infty}\sqrt[n]{\| A^n \|}$.
The positive kernel ${\cal Q}_\lambda (x,y)$ of the operator $Q_{\lambda}$ is the monotonically decreasing function of $\lambda >0$. Thus we have
$$
0 \ \leq \ Q_{\lambda_1} \ \leq \ Q_{\lambda_2}, \quad \mbox{ when } \quad \lambda_1 > \lambda_2,
$$
and consequently, the spectral radius $r( Q_{\lambda})$ is monotonically decreasing with respect to $\lambda >0$.
The convergence to 0 follows from (\ref{Q}).  \hfill     $\blacksquare$
\\


As follows from (\ref{5A}),  the equation on the eigenfunction $\psi$
\begin{equation}\label{L0}
(L_0 + V - \lambda) \psi \ = \ 0, \quad \lambda>0,
\end{equation}
can be written as
\begin{equation}\label{R}
Q_{\lambda} \psi (x) \  = \  \psi (x).
\end{equation}
where $Q_{\lambda}$ is a compact positive operator defined by (\ref{Q}).


Using  Lemma \ref{Lemma 2} we can conclude that if
\begin{equation}\label{lim2}
\lim_{\lambda \to 0+} r(Q_{\lambda}) \ > \ 1,
\end{equation}
then there exists such $\lambda>0$ that
\begin{equation}\label{r}
r( Q_{\lambda}) = 1, \quad \mbox{
and } \quad r( Q_{\lambda'}) < 1 \; \mbox{ for } \; \lambda'>
\lambda.
\end{equation}
If $r(Q_\lambda) <1$ for all $\lambda>0$, then the positive spectrum of $L$ is absent. For example, the positive spectrum of $L$ is absent in $ C_b({\mathbb{R}}^d)$ if $d \ge 3$, $V \in  L^1({\mathbb{R}}^d),\; 0 \le V(x) \le 1-\varepsilon <1$, and $L^1$-norm of $V$ is small enough.

From (\ref{r}) it follows by the Krein-Rutman theorem (\cite{KR}, Theorem 6.2) that 1 is the eigenvalue of $Q_\lambda$ with a positive eigenfunction $\psi_\lambda (x) >0$. Obviously, $\lambda$ is the maximal positive eigenvalue of the operator $L$, and $\psi_\lambda (x)$ is the ground state of the operator $L$. The uniqueness of the ground state $\psi_\lambda(x) >0$ of the operator $L$ in the space $L^2({\mathbb R}^d)$ follows from the positivity improving property of the semigroups $e^{tL_0}$ and $e^{tL}$, see e.g. \cite{RS4}, Theorem XIII.44. The last semigroup is positivity improving due to the Feynman-Kac formula.

\begin{lemma}
\label{Lemma 3} 1. If the ground state $\psi_\lambda(x) \in C_b({\mathbb{R}}^d)$, then $\psi_\lambda(x) \to 0$ as $|x| \to \infty$. \\
2. If the ground state $\psi_\lambda(x) \in L^2({\mathbb{R}}^d)$, then $\psi_\lambda(x) \in C_b({\mathbb{R}}^d) \cap  L^2({\mathbb{R}}^d)$ and  $\psi_\lambda(x) \to 0$ as $|x| \to \infty$. \\
3. If the ground state $\psi_\lambda(x) \in C_b({\mathbb{R}}^d)$ and  $V(x) \in  L^2({\mathbb{R}}^d)$, then
$\psi_\lambda(x) \in L^2({\mathbb{R}}^d)$. In particular, if the potential $V(x) \in C_0({\mathbb{R}}^d)$
 then   $\psi_\lambda(x) \in L^2({\mathbb{R}}^d)$.
 \end{lemma}

 {\it Proof.} 1. The ground state $\psi_\lambda(x)$ satisfies the equation (\ref{R}) with $Q_\lambda$ defined by (\ref{Q}).
Since  $V(x) \to 0$ as $|x| \to \infty$ and $G_\lambda(x) \in L^1({\mathbb{R}}^d)$ by (\ref{D}), then from
the Lebesgue convergence theorem it follows that the convolution of the functions $V \psi_\lambda$ and $G_\lambda$ tends to 0 for $|x| \to \infty$. Thus, all eigenfunctions $\psi_\lambda(x) \in C_b({\mathbb{R}}^d)$ for positive $\lambda$ tend to 0 when $|x| \to \infty$.

2. We again use the equation (\ref{R}) for the function $\psi_\lambda(x)$. Since $V(x)$ is bounded, then $(V \psi_\lambda)(x) \in L^2({\mathbb{R}}^d)$ and $(\widetilde{V \psi_\lambda})(p) \in L^2({\mathbb{R}}^d)$. Using (\ref{10A}) $\widetilde{G_\lambda}(p) \in L^2 ({\mathbb{R}}^d)$ ($\lambda>0$), we have $\widetilde{G_\lambda} (p) \widetilde{V \psi_\lambda}(p) \in L^1 ({\mathbb{R}}^d)$ as a product of two functions from $L^2({\mathbb{R}}^d)$. Consequently, $(A_\lambda V \psi_\lambda)(x) \in C_b({\mathbb{R}}^d)$, and $\psi_\lambda(x) = Q_\lambda \psi_\lambda(x) \in C_b({\mathbb{R}}^d)$, since $W_\lambda^{-1} \in C_b({\mathbb{R}}^d)$ for any $\lambda>0$.

3. The statement directly follows from (\ref{R}). \hfill     $\blacksquare$

\section{Ground state in $C_b({\mathbb R}^d)$}

Now we formulate conditions on $V(x)$ which give the existence of the ground state $\psi(x)$. We omit the subscript $\lambda$ in  $\psi_\lambda(x)$ in the subsequent text.

\begin{theorem}[small paradise]
\label{Theorem 2}  Assume that there exists $\delta >0$  such that
$V(x) = 1$ when $x \in B_\delta$, where $B_\delta$ is a ball of a radius $\delta$. Then the ground state of $L$ exists.
\end{theorem}

 {\it Proof.}  It is enough to show (\ref{lim2}).
Let us take for a given $0<\delta <1$ a "continuous approximation" of
the indicator function $\hat \chi_{B_\delta}(x)$:
$$
\hat \chi_{B_\delta}(x) = 0, \; x \in \mathbb{R}^d \backslash B_\delta, \quad
\hat \chi_{B_\delta}(x) = 1, \; x \in B_{0.9 \delta},
$$
where $B_{0.9 \delta} \subset B_\delta$, and  $0 \le \hat
\chi_{B_\delta}(x) \le 1, \ x \in B_\delta \backslash B_{0.9 \delta}$. Then for any
$\lambda \in (0,1)$ and any $ x \in B_\delta$ we get:
\begin{equation}\label{22A}
Q_{\lambda} \hat \chi _{B_{\delta}} (x) \ = \  \int_{B_\delta} {\cal
Q}_{\lambda} (x,y) \hat \chi _{B_{\delta}} (y) dy \ \ge \
\frac{1}{\lambda}  \int_{B_{0.9 \delta}} G_{\lambda} (x-y) dy \ \ge \
 \frac{ vol (B_{ 0.9 \delta})}{\lambda} \kappa_1,
\end{equation}
where $ vol (B_{ 0.9 \delta})$ is the volume of the ball $ B_{ 0.9 \delta} \subset B_\delta$, and $\kappa_1$ is defined as
\begin{equation}\label{k1}
\kappa_1 = \min_{x,y \in B_1} G_{1}(x-y) < \min_{\lambda \in (0, 1)} \min_{x,y \in B_\delta} G_{\lambda}(x-y)
\end{equation}

Thus $\lim_{\lambda \to 0+} r( Q_{\lambda}) = \infty$ for any $\delta >0$. \hfill     $\blacksquare$

\begin{remark}
\label{Remark 2A}
For any $\delta >0$ there is $\varepsilon>0$ such that if $V(x) \ge 1- \varepsilon$ for $x \in B_\delta$, then the ground state of $L$ exists.
\end{remark}

The proof of the statement follows the similar reasoning as above in Theorem \ref{Theorem 2}.

\begin{theorem}
\label{Theorem 3}  Assume that for some $\beta\in (0,1)$ there exists $R>0$ such that
\begin{equation}\label{Th3}
\beta \  \le \ V(x) \  \le \ 1, \quad x \in B_R.
\end{equation}
Then the ground state of the operator $L$ exists for $R=R(\beta)$ sufficiently large.
\end{theorem}

 {\it Proof.} We will prove that there exists a ground state $\psi$ of the operator $L$, such that $\psi \in L^2(\mathbb{R}^d)$. Then from Lemma \ref{Lemma 3} it follows that $\psi \in C_b(\mathbb{R}^d)$.

To prove the existence of $\psi \in L^2(\mathbb{R}^d)$ it is sufficient to verify that the quadratic form $(Lf,f)$ is positive for some $f \in L^2(\mathbb{R}^d)$. Let us take $f=\chi_{B_R}$. Then
\begin{equation}\label{3.1}
(L\chi_{B_R}, \chi_{B_R}) = (L_0 \chi_{B_R}, \chi_{B_R}) + (V \chi_{B_R}, \chi_{B_R}),
\end{equation}
and
\begin{equation}\label{3.2}
(V\chi_{B_R}, \chi_{B_R}) \ \ge \ \beta \ vol(B_R).
\end{equation}
For the operator $L_0$ we have:
$$
- (L_0 f,f) = \int \int a(y-x) (f(x) - f(y)) f(x) dy dx =
$$
$$
 \frac12  \int \int a(y-x) (f(x) - f(y))^2 dy dx.
$$
Consequently, the first term in (\ref{3.1}) can be written as
$$
- (L_0 \chi_{B_R},  \chi_{B_R}) = \int_{|x|<R} \int_{|y|>R} a(y-x)  dx dy  \le  \int_{|x|<R} \int_{|x|+|z| > R} a(z)  dx dz  =
$$
$$
C_d \int_0^R r^{d-1} \int_{|z| > R-r} a(z) dz dr = C_d \int_{\mathbb{R}^d} a(z)  \int_{(R-|z|)_+}^R r^{d-1} dr dz =
$$
$$
\frac{C_d R^d}{d} \int_{\mathbb{R}^d} a(z)  \left( 1- \left(1-\frac{|z|}{R} \right)^d_{+}  \right) dz.
$$
Here $\frac{C_d R^d}{d} = vol(B_R)$. Thus,
\begin{equation}\label{3.3}
\frac{1}{vol(B_R)} (L_0 \chi_{B_R},  \chi_{B_R}) \ \to \  0 \quad \mbox{ as } \quad R \to \infty.
\end{equation}
Finally, (\ref{3.1}) - (\ref{3.3}) imply that the operator $L$ has a positive discrete spectrum and a ground state if $R$ is taken sufficiently large.
\hfill     $\blacksquare$

\begin{theorem}
\label{Theorem 1}  Let $d=1,2$, and
 \begin{equation}\label{a3}
 \quad  \int_{\mathbb R^d} |x|^2 a(x) dx < \infty.
 \end{equation}

Then for any $V \not \equiv 0$ the ground state of $L$ exists.
\end{theorem}

 {\it Proof.} Conditions on the function $a(x)$ imply that $\tilde a
(p) = 1 - (Cp,p) + o(|p|^2)$ as $|p| \to 0$. Let us take a function $\varphi(x) \in C_0 ({\mathbb R}^d), \ \varphi \ge 0, \ \| \varphi \| =1$ such that  $(\widetilde{V \varphi}) (0) = \int V(x) \varphi(x) dx > 0$. Using the properties of the functions $a(x)$ and $V(x)$, see (\ref{a1})-(\ref{a4}), (\ref{6A}), we get that
$$
\tilde a(p), \, (\widetilde{V \varphi}) (p) \in L^2 ({\mathbb R}^d) \cap C_b ({\mathbb R}^d),
$$
and consequently, $\tilde a(p) (\widetilde{V \varphi}) (p) \in L^1 ({\mathbb R}^d) \cap C_b ({\mathbb R}^d)$.

Then for any $x \in supp \ \varphi$
$$
(Q_{\lambda} \varphi) (x) \ = \  \frac{1}{\lambda+1-V(x)} \int_{\mathbb{R}^d} G_\lambda (x-y) V(y) \varphi(y) \ dy \ =
$$
$$
\frac{1}{\lambda + 1 - V(x)} \int_{\mathbb{R}^d} \int_{\mathbb{R}^d} e^{-ipx+ipy} \frac{\tilde a(p) \ dp}{\lambda+1- \tilde a(p)} V(y) \varphi(y) \ dy \ =
$$
$$
\frac{1}{\lambda + 1 - V(x)} \int_{\mathbb{R}^d} \frac{  e^{-ipx} \ \tilde a(p) \  (\widetilde{V \varphi}) (p)}{\lambda + 1 - \tilde a(p)}  \ dp \to \infty \; \mbox{ as } \; \lambda \to 0.
$$

The continuous functions $U_\lambda(x) = (Q_\lambda \varphi)^{-1}(x)$ tend to 0 monotonically as $\lambda \to 0+$ by Remark \ref{Remark 1} and uniformly on $supp \ \varphi$ by the Dini theorem. Thus for any $c_0$ there exists $\lambda>0$ for which (\ref{c0}) is fulfilled.
Consequently, $\lim_{\lambda \to 0+} r(Q_{\lambda}) = \infty$ and (\ref{lim2}) holds. \hfill     $\blacksquare$



\section{Ground state in $L^2({\mathbb R}^d)$}

Now we can omit the condition of the continuity of $V(x)$. The statements of Lemmas \ref{Lemma 1}, \ref{Lemma 2} hold in the space  $L^2({\mathbb R}^d)$, since the operator $Q_\lambda$ defined by (\ref{Q}) is a compact positive operator in the order sense in $L^2({\mathbb R}^d)$. The further consideration of the problem in the space $L^2({\mathbb R}^d)$ is simpler because the operator $L=L_0 +V$ is a bounded self-adjoint operator in $L^2({\mathbb R}^d)$. The analysis of the operator $L$ is based on transformations of equation (\ref{L0}), that are analogous to the transformations exploited in the theory of Schr\"{o}dinger operators. As a result, the ground state problem can be reduced to the spectral analysis of a compact positive self-adjoint operator.

The equation (\ref{L0}) on the eigenfunction $\psi(x)$ can be rewritten in the following way:
\begin{equation}\label{V}
V^{1/2} (\lambda - L_0)^{-1} V^{1/2}  u \ = \ u,
\end{equation}
where$$u \ = \ V^{1/2} \psi, \quad \psi \ = \ (\lambda - L_0)^{-1} V^{1/2} u.$$
Inserting (\ref{mu}) to (\ref{V}) we get that (\ref{V}) is equivalent to the following equation:
\begin{equation}\label{tildeR}
 S_{\lambda} u \  = \  u, \quad \mbox{ where } \quad S_{\lambda} \  =  \ \frac{1}{\lambda+1}  W_{\lambda}^{-1} V^{1/2} A_{\lambda} V^{1/2},
\end{equation}
and $A_\lambda$ is defined by (\ref{a}). $S_{\lambda}$ is a compact positive operator, and it is similar to the compact positive symmetric operator
$$
\hat S_{\lambda} \  =  \  \frac{1}{\lambda+1}
W_{\lambda}^{-1/2}  V^{1/2} A_{\lambda} V^{1/2}  W_{\lambda}^{-1/2}.
$$
Consequently the spectra of operator $S_\lambda$ and self-adjoint operator $\hat S_\lambda$ are the same, and
\begin{equation}\label{rhatQ}
r(S_{\lambda}) \ = \ r(\hat S_{\lambda})  \  = \  \| \hat S_{\lambda}\|_{L^2}.
\end{equation}
The spectral radius $r(S_\lambda)$ of the operator $S_\lambda$ has the same properties as in Lemma \ref{Lemma 2}.

All statements of Theorems \ref{Theorem 2}, \ref{Theorem 3}, \ref{Theorem 1} are also valid in the space $L^2({\mathbb R}^d)$. In the proof of Theorem \ref{Theorem 1} we have to use the variational principle for $ \hat S_{\lambda}$ instead of Remark  \ref{Remark 2}. 


\begin{remark}
 If the potential $V \in L^1({\mathbb R}^d)$, then from (\ref{R}) and Lemma \ref{Lemma 3} it follows that $\psi(x) \in L^1({\mathbb R}^d)$ for any $\lambda>0$.
\end{remark}

\section{The subcritical regime}

We consider in this section a local perturbation in the form of "small paradise" for the subcritical regime: $m(x) \equiv m >1$. In this case the non local operator $L$ has a form
\begin{equation}\label{Lsub}
Lu(x) \ = \  - \tilde m(x) u(x) \ + \ \int_{\mathbb R^d} a(x-y) u(y) dy.
\end{equation}
It can be rewritten as
\begin{equation}\label{LD}
L u (x) \ = \ L_0 u (x) \ + \ D (x) u (x),
\end{equation}
where
\begin{equation}\label{L0D}
L_0 u(x) \ = \ \int_{\mathbb{R}^d} a(x-y) (u(y) - u(x)) dy,
\end{equation}
\begin{equation}\label{L0D1}
D(x) \ = \ 1- \tilde m(x) \ = \  \tilde V(x) - h, \quad
h= m-1>0.
\end{equation}
We assume that the function $\tilde V(x) \ = \ m - \tilde m(x), \; 0 \le \tilde V(x) \le m$ has a bounded support:
\begin{equation}\label{Vsub}
\tilde V(x) = 0 \;  \mbox{ as } \; |x| \ge R.
\end{equation}
If $\tilde V(x) \equiv 0$, then from (\ref{LD}) - (\ref{L0D1}) it follows that
$$
(Lu,u) \ \le \ -h(u,u), \quad h = m-1>0.
$$

The equation on a positive eigenvalue of the operator (\ref{LD}) and the corresponding eigenfunction $\psi(x)$
$$
L_0 \psi \ + \ D \psi \ = \ \hat\lambda \psi, \quad \hat\lambda>0,
$$
is equivalent to the above equations (\ref{L0}) - (\ref{R}) with
$$
{\cal Q}_{\lambda} (x,y) \ = \ \frac{ G_{\lambda}(x-y) \tilde V(y)}{ \lambda + 1- \tilde V(x)}, \quad 0 \le \tilde V(x) \le m
$$
under additional condition: $\lambda>h$.
Moreover, as follows from Lemma \ref{Lemma 3}, in this case the ground state $\psi (x) \in L^2({\mathbb R}^d)$.

\begin{theorem}[small paradise in the subcritical regime]
\label{Theorem 12}  For any $d~\ge~1$, assume that there exists $\delta >0$  such that
$\tilde V(x) = m$ when $x \in B_\delta$, where $B_\delta$ is a ball of a radius $\delta$. Then the ground state of $L=L_0+D$ exists.
\end{theorem}

{\it Proof.}  It is enough to prove inequality similar to (\ref{lim2}):
$$
\lim_{\lambda \to h+} r(Q_\lambda) >1.
$$

Using the same reasoning as in Theorem \ref{Theorem 2} for the normalized indicator function $\phi_\delta(x) = \frac{1}{\sqrt{|B_\delta|}} \chi_{B_\delta}(x)$ we get for any $ x \in B_\delta$:
$$
Q_{\lambda} \phi_\delta (x) \ = \  \int_{B_\delta} {\cal
Q}_{\lambda} (x,y) \phi_\delta (y) dy \ \ge \
\frac{m}{(\lambda+1-m) \ \sqrt{|B_\delta|}} \int_{B_{\delta}} G_{\lambda} (x-y) dy \ \to \ \infty
$$
as $\lambda \to h+$. Thus $\lim_{\lambda \to h+} r( Q_{\lambda}) = \infty$ for any $\delta >0$, and equation (\ref{R}) has a solution which is the ground state $\psi(x)>0$ corresponding to the maximal eigenvalue $\hat \lambda = \lambda - h>0$ of the operator (\ref{LD}). \hfill     $\blacksquare$
\\

Theorem \ref{Theorem 12} means that any local perturbation in the form of small paradise in the subcritical regime produces a crucial change in the asymptotic behavior of the system: instead of the exponentially decreasing population density one can find an exponentially increasing population everywhere in the space, and the density profile is described by the corresponding ground state $\psi(x)$.

\section{Concluding remarks}

Existence and uniqueness of the ground state of the operator $L$ in $L^2({\mathbb R}^d)$ immediately implies the following asymptotic formulas on the solution $u(x,t)$ of the evolution problem (\ref{PP}).

\begin{proposition}
 Assume that there exists a unique ground state $\psi>0$ of the operator L in $L^2({\mathbb R}^d)$,
and $\lambda>0$ be the maximal eigenvalue.

Then in $L^2({\mathbb R}^d)$ the following asymptotic formula holds:
\begin{equation}\label{T4}
 u(x,t) \ = \ e^{t \lambda} c_0 \psi(x) (1+ o(1)) \quad (t \to \infty),
\end{equation}
where $c_0 = (u_0, \psi)_{L^2}>0$ for any initial condition $u_0 \ge 0, \; u_0 \not \equiv 0$.

In $C_b({\mathbb R}^d)$ for any $u_0 (x) \ge 0, \ u_0 \not \equiv 0,$ and any bounded domain $D \subset \mathbb{R}^d$ we have
$$
\int_{D} u(x,t) dx \ \to \ \infty \quad \mbox{ as } \quad t \to
\infty.
$$
\end{proposition}

 {\it Proof.}  The operator (\ref{L}) is a bounded self-adjoint
operator in $L^2({\mathbb R}^d)$. Then  the asymptotic (\ref{T4}) is a direct
consequence of the spectral decomposition for the operator $L$.

Let us fix a ball $B$ such that $B \cap \ supp \ u_0 \neq \emptyset$, and denote $u_0^{B}(x) = u_0(x) \cdot \chi_{B}(x)$. Then $u_0^{B}(x)
\in L^2({\mathbb R}^d), \ u_0^B \not \equiv 0$. Using the positivity of the semigroup $e^{tL}$
we have for any bounded domain $D$
$$
\int_{D} u(x,t) dx =  \langle u(\cdot,t),  \chi_{D}(\cdot) \rangle = \langle
e^{tL} u_0, \chi_{D} \rangle \ge
$$
\begin{equation}\label{27}
\langle e^{tL} u_0^B, \chi_{D} \rangle  \sim  e^{\lambda t} c_0
(\psi, \chi_{D})_{L^2} \to \infty \quad (t\to \infty)
\end{equation}
with $c_0 = (u_0^B, \psi)_{L^2}>0$.
\hfill     $\blacksquare$

{\bf Conclusions.} Since $u(t,x), x \in \mathbb{R}^d$ describes the density
of the population at time $t$, then the asymptotic (\ref{27}) means
that, in the case when the operator $L$ has a positive eigenvalue
$\lambda>0$, the population is exponentially increasing everywhere in the space.
Moreover in the case $L^2({\mathbb R}^d)$, for any initial density $u(0,x) =
u_0 (x) \ge 0$ the shape of $u(t,x)$ tends to the shape of
the ground state $\psi(x)$ of the operator $L$ up to the multiplication on $e^{\lambda t}$.

If $m(x) \equiv 0$, then the total mass grows as $e^t$. Thus if the positive eigenvalue exists, then the exponent $\lambda$ of the growth is in the interval $(0,1]$.

The ground state of the operator $L$ appears in two cases: \\ 1) if there is a bounded domain of a positive volume,
where the mortality is equal to 0 (Theorem \ref{Theorem 2})  (small paradise); the existence of the ground state doesn't depend on the volume of this domain; \\
2) if the mortality is less than 1 in a bounded domain, where the size of the domain depends on the bounds of the mortality (Theorem \ref{Theorem 3}).

To observe the exponential growing of the density in small dimensions $d=1,2$ with a fast decaying dispersal kernel it is enough to
have any small region where mortality is less than 1 (Theorem \ref{Theorem 1}).

We found that perturbations in the form of small paradise are very powerful: they switch a subcritical regime in the system (with exponentially decreasing population density) to a supercritical regime with exponentially increasing population (Theorem \ref{Theorem 12}).

Next natural question is what happens if the mortality $m(x)$ is greater than 1 (or even $m(x)$ is a growing function) outside of a region, where the mortality has a local negative fluctuations. Is it possible that an active growing inside of a bounded region can be stronger than the influence of a large mortality outside? We suppose to study this question in a forthcoming paper. We also plan to continue spectral analysis of the operator $L$ in more details including the study of continuous spectrum and structure of the ground state.


\begin{thebibliography}{20}

\bibitem{BC} H. Berestycki, J. Coville, H.-H. Vo, On the definition and the properties of the principal eigenvalue of some nonlocal operators, http://arxiv.org/pdf/1512.06529v1.pdf (2015)

\bibitem{Cov} J. Coville, On a simple criterion for the existence of a principal eigenfunction of some nonlocal operators, J. Differential Equations 249 (2010), 2921-2953.

\bibitem{DS} N. Dunford, J. Schwartz, Linear Operators, Wiley, NY 1988.

\bibitem{KKP} Yu. Kondratiev, O. Kutoviy, S. Pirogov, Correlation functions
and invariant measures in continuous contact model, Ininite
Dimensional Analysis, Quantum Probability and Related Topics Vol.
11, No. 2, 231-258 (2008)


\bibitem{KS} Yu. G. Kondratiev and A. Skorokhod, On contact processes in continuum,
Ininite Dimensional Analysis, Quantum Probability and Related Topics
Vol. 9, 187-198 (2006)

\bibitem{KR} Krein, M.G.; Rutman, M.A., Linear operators leaving
invariant a cone in a Banach space, Uspehi Matem. Nauk (in Russian)
3, p. 3-95 (1948). English translation: Krein, M.G.; Rutman, M.A.,
Linear operators leaving invariant a cone in a Banach space, Amer.
Math. Soc. Translation 26 (1950).


\bibitem{RS1} M. Reed, B. Simon, Methods of modern mathematical physics, Vol.1, Academic Press, NY 1972

\bibitem{RS4} M. Reed, B. Simon, Methods of modern mathematical physics, Vol.4, Academic Press, NY 1978



\end{thebibliography}
\end{document}